\documentclass[amsmath,amssymb,preprint,nofootinbib]{revtex4}
\usepackage{amsfonts} 
\usepackage{graphicx} 
\usepackage{epsfig}
\usepackage{bm}
\everymath{\displaystyle}
\def \a {{\tilde{\alpha}\;}}
\def \as {{\tilde{\alpha}_s\;}}
\def \S {S}
\def \Suneq {S_{\rm{uneq}}}
\def \Xuneq {X_{\rm{uneq}}}
\def \SeqS {S_{{\rm{eq,s}}}}
\def \XeqS {X_{{\rm{eq,s}}}}
\def \SuneqS {S_{{\rm{uneq,s}}}}
\def \XuneqS {X_{{\rm{uneq,s}}}}

\def \vr {|{\bf{v}}|}

\begin{document}
\title{Relativistic Coulomb  $\boldsymbol{S}$-factor of Two Spinor Particles with Arbitrary Masses}

\author{Yu.~D. Chernichenko}
\email{chern@gstu.by}
\affiliation{Gomel State Technical University, Gomel, 246746, Belarus}

\author{L.~P. Kaptari}
\email{kaptari@theor.jinr.ru} \affiliation{Bogoliubov Laboratory of
Theoretical Physics, Joint Institute for Nuclear Research, Dubna, 141980, Russia}

\author{O.~P. Solovtsova}
\email{olsol@theor.jinr.ru}
 \affiliation{International Center for Advanced Studies,
Gomel State Technical University,  246746,  Belarus}
\affiliation{Bogoliubov Laboratory of
Theoretical Physics, Joint Institute for Nuclear Research, Dubna, 141980, Russia}

\begin{abstract}
A new   resummation of the $S$-factor of a composite system of two relativistic spin-1/2 particles of arbitrary masses  interacting via a Coulomb-like chromodynamical potential is presented. The analysis is performed in the framework of a relativistic quasipotential approach based on  the Hamiltonian formulation of  the  covariant  quantum field theory in the  relativistic configuration representation. The pseudoscalar, vector, and pseudovector systems are considered and the behaviour of the $S$-factor near the threshold and  in the relativistic limit is investigated in detail. The spin dependence  of  the $S$-factors is discussed as well. It is argued that at the threshold  the contribution of spins  significantly reduces the Sommerfeld effect, while at   ultrarelativistic velocities their  role   diminishes  and the $S$-factor becomes basically the same as for the spinless systems.  A connection between the new   and  previously obtained  S-factors for spinless  particles of arbitrary masses and for relativistic spinor particles of equal masses is established.
\end{abstract}


\maketitle

\vspace{-1.0cm}

\section{Introduction}

It is well known from quantum electrodynamics (QED) that the so-called
Coulomb resummation factor plays an important role in describing the
system of two charged particles near the threshold \cite{Schwinger70}.
The resummation performed on the basis of the nonrelativistic Schr\"odinger equation with the Coulomb potential,
$V(r)=-\alpha/r$, where $\alpha$ is the fine structure constant, leads to the Sommerfeld--Gamov--Sakharov
factor~\cite{Sommerfeld,Gamov:28,Sakharov} which, for two particles of equal masses, reads as:
\begin{equation} \label{SS-factor}
S_{\rm nr}\,=\,\frac{X_{\rm nr}}{1\,-\,\exp(-X_{\rm nr})}\,,\quad X_{\rm nr}\,=\,\frac{\pi\alpha}{v_{\rm nr}},
\end{equation}
where $v_{\rm nr }$ denotes the velocity of each particle in the
center-of-mass frame.
At small values of $\;{\alpha/v_{\rm nr}}$, the $S$-factor
(\ref{SS-factor}) behaves as $S_{\rm nr}\,\sim 1+\pi \alpha/(2v_{\rm
nr})$, which implies that at low velocities the corresponding cross
sections increase. As originally mentioned by
Sommerfeld~\cite{Sommerfeld}, this enhancement is directly related to
the wave function of the relative motion in the Coulomb field
evaluated at the origin, $|\psi(0)|^2$, i.e. to the probability to
find two interacting particles close to each other. Such an
increase of the $S$-factor is often referred to as the Sommerfeld
effect. A similar phenomenon was also predicted  and then confirmed 
by Gammov~\cite{Gamov:28} in nuclear reactions when interacting
particles overcome the Coulomb barrier and by Sakharov~\cite{Sakharov} in
electron-positron pair production. Intensive studies of the
Sommerfeld-effect demonstrate the crucial role it plays in
understanding   a large variety of inelastic processes with heavy
particles moving slowly in a thermal environment, see, e.g.,
Refs.~\cite{Hisano:2003ec,Cassel:2009wt,Kim:2019qix,Mahbubani:2019shd,Dremin:2020vdb,ArkaniHamed:2008qn,Binder:2019erp}.
These processes are encountered in many physical situations, and
nowadays there is renewed  interest in detailed investigations of the
Sommerfeld effect. A classic example is given by astrophysical
nuclear reactions taking place within the electromagnetic plasma of
stars. In particle physics, one may mention  heavy Dark Matter
particles produced by pair annihilation in the early Universe. If  
Dark Matter in the Universe contains interactions with the Standard Model
particles, the  annihilation may give visible imprints in the
cosmic rays~\cite{Hisano:2003ec}.
Within quantum chromodynamics (QCD), an increase in the $S$-factor may
occur in heavy quark and anti-quark pair annihilation into light
quarks and gluons within a quark-gluon plasma environment generated in
heavy ion collision experiments. Bound states contribute to the
annihilation process, and enhancement factors of up to $\sim 100$ can
be encountered, see, e.g., Ref.~\cite{Kim:2019qix}. Other examples
where the Sommerfeld effect is crucial, include threshold production
of heavy states at colliders  and partial decay rates when the
products have large phase space suppression. The Sommerfeld effect in
boosting the dark matter signal with Coulomb resonances has been
considered, to some extent, in
Ref.~\cite{Mahbubani:2019shd,Cassel:2009wt}, where the $S$-factor was
computed within a nonrelativistic approach for attractive and
repulsive Coulomb and Yukawa (between neutral particles)
potentials. Details of the theoretical consideration of the Sommerfeld
enhancement in   Dark Matter can be found in, e.g.,
Refs.~\cite{ArkaniHamed:2008qn,Binder:2019erp}.

Another manifestation of the enhancement of the
Sommerfeld-Gamow-Sakharov factor can be clearly seen in abundant
creations of di-lepton pairs with low masses in heavy-ion
collisions~\cite{Dremin:2020vdb}, which is directly related to the
physics at the NICA and LHC colliders. It should be noted that most of the 
theoretical studies of the $S$-factor have been performed within nonrelativistic approaches originating from the Schr\"odinger equation.
Coming back to Eq.~(\ref{SS-factor})  one shall stress that, in spite
of the expansion of the $S$-factor in a power series $(\alpha/v)^n$
with respect to the coupling constant $\alpha$ reproduces the
threshold singularities of the Feynman diagrams,  in the threshold-near
region, $v\to 0$, the parameter $\alpha/v$ can no  longer be adequate
to cut off the perturbative series. Consequently, the $S$-factor
should be taken into account in its entirety.

A similar situation also arises also in QCD because in describing a
quark-antiquark system near the threshold,
$s_{th}=(m_q+m_{\bar{q}})^2$, the expansion parameter $\alpha_s/v$
becomes singular when $v\to 0$~\cite{App-Politzer75,PQW}. Here
$\alpha_s$ denotes the strong coupling constant, and the quark velocity
$v$ for the case of equal masses, $m_q = m_{\bar{q}}\equiv m$, reads as
\begin{equation} \label{v-velocity}
v=\sqrt{1-\frac{4m^2}{s}}\,,
\end{equation}
where $\sqrt{s}$ is the total energy of the considered two particles in their c.m. system.
Therefore, in order to obtain a meaningful result,  threshold
singularities of the form $(\alpha_s/v)^n$ should be summed, which
can be achieved by solving the corresponding relativistic problem.
It is important to note that the problem of accounting for the
relativistic effects in a few body system  is a more general and
longstanding task. Up to now there are no reliable relativistic
equations derived from the first principles to describe few body
bound and/or continuum states near the threshold. For this reason, one
is forced to use phenomenological or quasi-phenomenological formalisms
to relativistically describe these systems.
One can mention several approaches  defining the relativistic wave functions of  few-body systems. One of the
approaches is based on the fully covariant and Lorentz-invariant Bethe-Salpeter (BS) formalism~\cite{BS}.
As is well known, in Minkowski space the BS equation is singular, hence in order to avoid uncertainties in
numerical calculations, one usually performs the Wick rotation of the equation to Euclidean space and solves the BS
equation along the imaginary energy axis. Such a representation is extremely useful for covariant calculations of
diagonal matrix elements which are the same in both  Euclidean and Minkowski spaces. For instance, within the BS
formalism one can relativistically describe  the main properties of   few-body bound states  such as
glueballs~\cite{LPK,Fischer}, quark-antiquark systems (mesons)~\cite{Drkin1,Dorkin2,ALkofer}, two-nucleon
bound state (deuteron)~\cite{dorkin3}, etc. However, not all physical observables can be obtained from
calculations in Euclidean space by the inverse Wick rotation back to Minkowski space. Consequently,
one needs to solve the BS equation directly in Minkowski space by finding new mathematical methods
allowing one to overcome the difficulties resulting from the singularities of the BS equation, c.f.~\cite{Ydrefors,Karmanov}. Also, within the BS formalism one encounters difficulties related to the
problem of probabilistic interpretation of the BS amplitudes and BS vertex functions. This problem can be partially
solved by considering different approximations to the BS equation like the Gross Covariant Spectator
Equation~\cite{CSE}.

Another strategy of relativization of calculations consists in elaborating relativistic analogues of the usual
three-dimensional Schr\"odinger equation. These approaches are mainly based on the 1949 seminal paper~\cite{Dirac},
where P.A.M. Dirac proposed several peculiar representations of the Poincar\'{e} group, in strict relation with the
choice of possible space-time hyper surfaces without a time-like direction. Among them the most popular is the approach
based on the Light Front dynamics which allows for a specific description of the dynamics of relativistic interacting
systems~\cite{LF,Pace}.

Eventually, a tempting and rather popular approach is the one based on the quasipotential (QP) equations in
Lobachevsky space~\cite{LogunovTav,Kadyshevsky68,KadMS:72,Faustov73}. The QP equations
are differential ones with the structure very similar to the Schr\"odinger equation; however, the
interaction quasipotential is now energy-dependent. Note that  at vanishing curvature the solution
of the QP equation reduces to the known nonrelativistic approaches. Albeit the QP wave functions, being
three-dimensional, admit a probabilistic interpretation, they at the same time have the main advantages
(renormalizability, analyticity, etc.) of the completely covariant field theory, see e.g. Ref.~\cite{KadMS:72}.

For the first time, the relativization of the $S$-factor
(\ref{SS-factor}) in the case of two particles with equal masses was
obtained in Refs.~\cite{Fadin88,Fadin95}, and the result is reduced to the
replacement of  the nonrelativistic velocity
$v_{\rm nr}$ in Eq.~(\ref{SS-factor}) by the relativistic one, $v_{\rm nr}\rightarrow\, v$.
Exactly the same form of the $S$-factor but with the change
$v_{\rm{nr}}\rightarrow\, v/(1+v^2)$ was later on suggested in
Ref.~\cite{Hoang97}.
Note here that the relativistic generalization of the $S$-factor is
obviously not unique, since there are numerous ways of expressing the
nonrelativistic velocity in terms of the invariant energy $\sqrt{s}$.  
Another form of the relativistic generalization of the $S$-factor was obtained
in~Ref.~\cite{YoonWong:0005}.
The relativistic $S$-factor for two particles of arbitrary masses was 
for the first time  presented in Ref.~\cite{Arbuzov94} (see also
Refs.~\cite{Arbuzov2011,Arbuzov2019}). It was considered within the framework of a version of the relativistic quantum mechanics, on the basis of the Schr\"odinger equation for the wave function $\Psi(t,
{\boldsymbol{x_1}},{\boldsymbol{x_2}})$  in a specific frame of reference,
${\boldsymbol{p}}_2=-{\boldsymbol{p}}_1m_2/m_1$, and treating
${\boldsymbol{p}}_{1,2}$ as differential operators.

In the present paper, we employ the relativistic quasipotential (RQP)
approach~\cite{Kadyshevsky68,KadMS:68,KadMM:70,KadMS:72},
which, in our opinion, is the most suitable one for the goal of
investigations of the effects of relativisation of the $S$-factor for
systems consisting of two relativistic, spin 1/2, particles with
arbitrary masses $m_1$ and $m_2$. The pseudoscalar, vector, and
pseudovector systems are considered, and the behavior of the
corresponding $S$-factors is analyzed in the nonrelativistic,
relativistic and ultrarelativistic limits.

Our paper is organized as follows. In the next section, we present the
$S$-factor calculated within the RQP approach for two spinless
relativistic particles. Explicit, analytical expressions for the $S$-factor of two spinor particles
with equal and arbitrary masses are presented in Secs.~III and IV,
respectively. The results of numerical calculations  for pseudo-scalar, vector and pseudo-vector systems are presented in Secion V. The conclusions and summary are collected in Secs.~VI and VII.

\section{Relativistic $\boldsymbol{S}$-factor for spinless particles}

In this section we consider the $S$-factor corresponding to systems of two relativistic
spinless particles with equal masses $m_1=m_2=m$.
It is important to note that the use of the RQP approach for our task is justified
by   that the BS amplitude $\Phi_{\rm BS}(x),x=(x_0,\bf{x})$,
that parameterizes, e.g., the Drell ratio $R(s)$ in QCD (see
also Ref.~\cite{Adel-Yndurain:95}), is evaluated at $x=0$~\cite{BarbieriCR73}.
Consequently, the BS amplitude can be related to the RQP wave function in momentum space,
$\Psi_{\chi}({\bf p})$, and in the configuration representation, $\psi_{\chi}({\bf r})$, as
\begin{eqnarray}
\Phi_{\rm BS}(x=0)=\frac{1}{(2\pi)^3}\int d\Omega_{\bf p}\Psi_{\chi}(\mathbf{p})
=\psi_{\chi}({\bf r})\Big|_{r=i\lambda},\label{BS-Psi-psi-RQP}
\end{eqnarray}
where $\chi$ is the rapidity related to the total c.m. energy of particles $\sqrt{s}$ as $\sqrt{s}=2m\cosh\chi$, $\lambda=1/m$ is the Compton wavelength of the particle of mass $m$, $d\Omega_{\bf p}=(m d{\mathbf{p}})/p_0$
is the invariant space volume in the Lobachevsky space realized on the hyperboloid $p_0^2-{\mathbf{p}}^2=
q_0^2-{\mathbf{q}}^2=m^2$. According to Eq.~(\ref{BS-Psi-psi-RQP}), the $S$-factor within the RQP approach
is defined in terms of the corresponding wave function in the continuum, $\psi_{\chi}({\bf r})$, as
follows~\cite{Milton-Solovtsov_ModPL:01}
\begin{eqnarray}
\label{determination-S-factor-eq}
S_{\rm\,RQP}(\chi)=\lim_{r\to{i\lambda}}\left|\psi_{\chi}({\bf r})\right|^2.
\end{eqnarray}
The nonrelativistic replacement of the amplitude $\chi_{\rm BS}(x)$ by the wave function, which obeys the
Schr\"odinger equation with the Coulomb potential, leads to formula (\ref{SS-factor}).

Based on the RQP approach, Milton and Solovtsov obtained the following expression~\cite{Milton-Solovtsov_ModPL:01}:
\begin{equation}
\label{S-rel-eq} \S(\chi)=\frac{X(\chi)}{1-\exp\left[-X(\chi)\right]}\,,\,~X(\chi)=\frac{\pi\a}{\sinh\chi},
\end{equation}
where
$\a$ denotes the coupling constant in the Coulomb-like potential $V(r)=-\tilde{\alpha}/r$. The use of
Eq.~(\ref{S-rel-eq}) in QCD requires the replacement of $\a \rightarrow 4\alpha_{s}/3$, where ${4}/{3}$ is due to the
SU(3) color factor. Here, it is worth emphasizing that in the above mentioned method~\cite{Milton-Solovtsov_ModPL:01},
the possibility of transformation of the RQP equation from the momentum space to the relativistic configurational
representation in the case of two particles of equal masses~\cite{KadMS:68} has been widely employed. Note also that the Coulomb quasipotential   formally has the same form as the nonrealistic Coulomb potential since its behavior in the momentum Lobachevsky space corresponds to the static quark-antiquark potential $V_{q {\bar q}}\sim{\bar\alpha_s}(Q^2)/Q^2$ for which ${\bar\alpha_s}(Q^2)\sim 1/\ln Q^2$ plays a role of effective
charge~\cite{SavrinS80}. Thereby, one accumulates a dominant effects induced by the running QCD coupling.

The function $X(\chi)$ in Eq.~(\ref{S-rel-eq}) can be expressed in
terms of the velocity $v$ defined by Eq.~(\ref{v-velocity}) as
$X(v)=\pi\tilde{\alpha}\sqrt{1-v^2}/v$. It becomes obvious that the
shape of the $S$-factor in the relativistic case remains the same as in
Eq.~(\ref{SS-factor}), albeit with a replacement
$$v_{\rm{nr}}\rightarrow\,\frac{v}{\sqrt{1-v^2}}\,.$$
The presence of the square root ${\sqrt{1-v^2}}\,$ in the denominator
is essential since it provides the correct expected relativistic limit
($v\rightarrow 1$) of the $S$-factor in
Eq.~(\ref{S-rel-eq})~\cite{Milton-Solovtsov_ModPL:01,SOP-Ch10}. Notice
that the relativistic limits of the $S$-factors from
Refs.~\cite{Hoang97,YoonWong:0005} differ substantially from the
relativistic limit ($v\to1$) of the $S$-factor corresponding to
Eq.~(\ref{S-rel-eq}). A relativistic Coulomb-like resummation factor
for   arbitrary masses and orbital moment $\ell\geq 1$, called the $L$-factor,
was investigated in Ref.~\cite{SOP-ChTMF-11}.

Applications of the relativistic $S$-factor~(\ref{S-rel-eq}) in
describing some hadronic processes can be found in
Refs.~\cite{MSS_Adler:01,SS_NonLin:02,MSS:06}. Also, the
factor~(\ref{S-rel-eq}) was applied in Ref.~\cite{Milton:08} to
reanalyze the mass limits obtained for magnetic monopoles, which might
have been produced at the Fermilab Tevatron.

Generalization of the relativistic $S$-factor (\ref{S-rel-eq}) to
the case of two relativistic particles of unequal masses $m_1$ and $m_2$
can be written in terms of the velocity  $u$  as ( for details see Refs.~\cite{SOP-Ch10,SOP-ChTMF-11})
\begin{eqnarray}
\label{S-rel-uneq}
\Suneq(u)=\frac{X_{\rm{uneq}}(u)}{1-\exp\left[-X_{\rm{uneq}}(u)\right]}; \quad 
\Xuneq(u)=\frac{\pi\tilde{\alpha}\sqrt{1-u^2}}{u},
\end{eqnarray}
where the subscript ``uneq'' indicates the  quantities related to the case of unequal masses, 
and the velocity $u$ is determined by the expression
\begin{equation}
\label{u-uneq} u=\sqrt{1-\frac{4{m'}^2}{s-(m_1-m_2)^2}}.
\end{equation}
Here $m'=\sqrt{m_1m_2}$ is the mass of an effective   particle with   3-momentum
${\bf\Delta}_{k',m'\lambda_{\mathcal{Q}}}$    and energy  $\Delta_{k',m'\lambda_{\mathcal{Q}}}^0$
   proportional to the total c.m. energy of particles, $\sqrt{s}$, see Refs.~\cite{KadMM:70,KadMS:72}:

\begin{eqnarray}
\label{s-uneq-Delta-k'} && \sqrt{s}=\sqrt{(k_1+k_2)^2}=
\frac{m'}{\mu}\Delta_{k',m'\lambda_{\mathcal{Q}}}^0 \, ,\\
&& \Delta_{k',m'\lambda_{\mathcal{Q}}}^0= \sqrt{m'^2+{\bf\Delta}_{k',m'\lambda_{\mathcal{Q}}}^2},\notag
\end{eqnarray}
where $\mu=m_1m_2/(m_1+m_2)$ is the reduced mass of a composite
particle with 4-momentum $\mathcal{Q}=q_1+q_2$ and total mass
$M_{\mathcal{Q}}=\sqrt{\mathcal{Q}^2}$; $\Delta_{k',m'\lambda_{\mathcal{Q}}}^0$ and
${\bf\Delta}_{k',m'\lambda_{\mathcal{Q}}}$ are the time
and spatial components of the 4-vector $\Lambda^{-1}_{\lambda_{\mathcal{Q}}}k'=
\Delta_{k',m'\lambda_{\mathcal{Q}}}$ of the effective relativistic
particle in Lobachevsky space corresponding to the Lorentz boost
$L=\Lambda^{-1}_{\lambda_{\mathcal{Q}}}$, where $\lambda_{\mathcal{Q}}$ is the
4-velocity vector of the composite system, $\lambda_{\mathcal{Q}}=
(\lambda_{\mathcal{Q}}^0;\boldsymbol{\lambda}_{\mathcal{Q}})=
\mathcal{Q}/\sqrt{\mathcal{Q}^2}$ (see Refs.~\cite{KadMM:70,KadMS:72,ChernYAF2014} for details):
\begin{eqnarray}
\label{Delta-k'} && \Delta_{k',m'\lambda_{\mathcal{Q}}}^0=(\Lambda^{-1}_{\lambda_{\mathcal{Q}}}k')^0=
k_{0}'\lambda_{\mathcal{Q}}^0-{\bf k}'\cdot{\boldsymbol{\lambda}_{\mathcal{Q}}}\\
&& =\sqrt{m'^2+{\bf\Delta}_{k',m'\lambda_{\mathcal{Q}}}^2},\notag\\[0.2cm]
&&{\bf\Delta}_{k',m'\lambda_{\mathcal{Q}}}=\Lambda^{-1}_{\lambda_{\mathcal{Q}}}{\bf k}'=
{\bf k}'(-)m'{\boldsymbol{\lambda}_{\mathcal{Q}}}\notag\\
&& ={\bf k}'-{\boldsymbol{\lambda}_{\mathcal{Q}}}\left(k_0'- \frac{{\bf k}'\cdot{\boldsymbol{\lambda}_{\mathcal{Q}}}}
{1+\lambda_{\mathcal{Q}}^0}\right).\notag
\end{eqnarray}

In the case when $m_1=m_2\equiv m$, the velocity (\ref{u-uneq}) of the effective particle
reduces to the velocity $v$, Eq.~(\ref{v-velocity}). The 3-momentum ${\bf\,k'}$ of the
effective particle in Lobachevsky space is related to the relative relativistic velocity
${\bf\,v}$ of the composite system by the following expression~\cite{KadMS:72}
\begin{equation}
\label{k'-velocity-v-rel}{\bf\,k'}^2=2\mu^2\left(\frac{1}{\sqrt{1-{\bf\,v}^2}}-1\right).
\end{equation}

Recall that  within the RQP approach \cite{KadMM:70,KadMS:72}  the
modulus of the relative relativistic velocity $\vr$ of two particles can be
expressed in terms of their total c.m. energy $\sqrt{s}$ (see
Ref.~\cite{SOP-Ch10}) as
\begin{equation}
\label{v-rel} \vr=\displaystyle2\sqrt{\frac{s-(m_1+m_2)^2}{s-(m_1-m_2)^2}}\left[1+\frac{s-(m_1+m_2)^2}
{s-(m_1-m_2)^2}\right]^{-1},
\end{equation}
that is exactly the same as in, e.g., Ref.~\cite{Arbuzov94}.
From Eqs.~(\ref{u-uneq}) and (\ref{v-rel}) one can infer that
\begin{equation}
\vr=\frac{2u}{1+u^2}. \label{otn}
\end{equation}
It is obvious that the relativistic limit ($u\to1$) of the $S$-factor
considered in Ref.~\cite{Arbuzov94} \footnote{In Ref.~\cite{Arbuzov94}
the notation $T(|\bf v|)$ is used instead of $S(|\bf v|)$ adopted in
the present paper.}
\begin{equation}
S(\vr)=\frac{2\pi\alpha/{\vr}}{1-\exp\left(-2\pi\alpha/\vr \right)}\,,\label{S-factor-A}
\end{equation}
differs from the   $S$-factor defined by Eq.~(\ref{S-rel-uneq}) for which the relativistic limit is
$S_{\rm uneq}(u)\overset{{ u\to 1 }}{\,\,\xrightarrow{\hspace*{0.4cm}}1}$.
However, for small values of $u$ the $S$-factors defined by
Eqs.~(\ref{S-rel-uneq}) and Eq.~(\ref{S-factor-A}) provide the same
result.

\section{$\boldsymbol{S}$-factor for spinors of equal masses}
For the case of two spin-1/2 particles interacting via
a Coulomb-like potential in $s$-state ($\ell=0$), the wave function in the configuration representation
 and the corresponding  $S$-factor were thoroughly investigated
within the  RQP approach in Refs.~\cite{skachkov,ChYAF2019,CKS-2020}. Furthermore,   a closed analytical expression for the $S$-factor has been found
\begin{equation}
\label{S-factor-spin} \SeqS(\chi)=\frac{\XeqS (\chi)}
{1-\exp \left[-\XeqS (\chi)\right]}
e^{-\pi\tilde{\rho}}\left|\Gamma(2+i\tilde{\rho})
{_2}F_1(1+iB,-i\tilde{\rho};2;1-e^{-2\chi})\right|^2,
\end{equation}
where $\Gamma(2+i\tilde{\rho})$ is the familiar Euler gamma function,
${_2}F_1(1+iB,-i\tilde{\rho};2;1-e^{-2\chi})$ is the hypergeometric
function and the subscript ``s'' refers to spinors. In
Eq.~(\ref{S-factor-spin}) the quantities $\XeqS(\chi)\;$ and
$\tilde{\rho},\,B$ are defined as
\begin{eqnarray*}
&&\XeqS(\chi)=\frac{\pi\tilde{\alpha}(a\cosh^2\chi+b)}{2\sinh\chi},\\[0.2cm]
&& \tilde{\rho}=\frac{\tilde{\alpha}a\cosh\chi}{4},\,B=\frac{\tilde{\alpha}(a\cosh^2\chi+b)}{4\sinh\chi},
\end{eqnarray*}
where  the   parameters $a$ and $b$ for different total spin of the system are
\begin{eqnarray}
\label{parameter-a-b-eq-spin} a=\left\{\begin{array}{ll}
\displaystyle\,\,\,\,\,1 & \text{\rm for}\,\hat{O}=\gamma_5\,\text{(\rm pseudoscalar)},\\[0.1cm]
\displaystyle\,\,\,\,\frac{1}{2} & \text{\rm for}\,\hat{O}=\gamma_\mu\,\text{(\rm vector)},\\[0.3cm]
\displaystyle-\frac{1}{2} & \text{\rm for}\,\hat{O}=\gamma_5\gamma_\mu\,\text{(\rm pseudovector)};
\end{array}\right.\\[0.4cm]
\hspace{-0.9cm} b=\left\{\begin{array}{ll}
\displaystyle0 & \text{\rm for}\,\hat{O}=\gamma_5\,\text{(\rm pseudoscalar)},\\[0.3cm]
\displaystyle\frac{1}{4} & \text{\rm for}\,\hat{O}=\gamma_\mu\,\text{(\rm vector)},\\[0.3cm]
\displaystyle\frac{3}{4} & \text{\rm for}\,\hat{O}=\gamma_5\gamma_\mu\,\text{(\rm pseudovector)}.
\end{array}\right.\label{parameterbprime}
\end{eqnarray}

It can be shown that the relativistic limit, $v \to 1$ ($\chi\gg 1$), of the $S$-factor
is
\begin{equation} \label{lim-rel-S-factor-spin}
\SeqS (\chi)|_{\chi\gg 1}\simeq
\frac{2\pi(B-\tilde\rho)}{1-\exp[-2\pi(B-\tilde\rho)]}.
\end{equation}
which is valid for all values of the spin parameters $a$ and $b$.

\section{$\boldsymbol{S}$-factor for two spinors
of arbitrary masses}

\subsection{Relativistic Coulomb wave function}

\begin{figure}[htb]
\centering\includegraphics[width=0.5\textwidth]{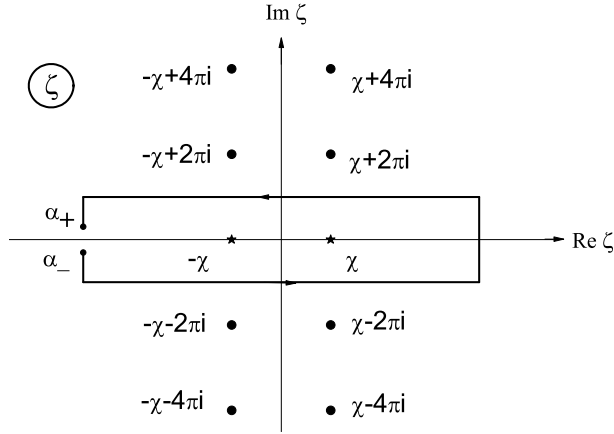}
\vspace{0.2cm}
\caption{Integration contour in
Eq.~(\ref{form-solution-varphi-ell=0-int-uneq-rho-spin}) and singularities of the
function~(\ref{solution-R-0-uneq-spin}) in the complex $\zeta$-plane.} \label{Fig-zeta-pl2}
\end{figure}

In this subsection we consider, in some detail, the wave function of
two relativistic spinor particles with arbitrary masses $m_1$ and $m_2$
interacting via a Coulomb\--like chromodynamical potential
\begin{equation} \label{Coulomb}
V(\rho)=-\frac{\as}{\rho}\,,
\end{equation}
where $\as\equiv{4\alpha_s}/{3}>0$ and $\rho=rm'$.
Correspondingly, the RQP equation for the radial wave function, $\varphi_{0}(\rho,\chi')$, acquires the form
\begin{eqnarray}
\label{eqvarphi-ell=0-int-uneq-rho-spin} && \hspace{-0.3cm}
\int\limits_{0}^{\infty}{d\chi}\Big[\left(\cosh\chi'-\cosh\chi\right)
\sin\rho\chi\int\limits_{0}^{\infty}d{\rho'}\sin(\rho'\chi)\varphi_{0}(\rho',\chi')\Big]\nonumber\\[-0.2cm]
&&\hspace{-0.8cm} \\ [-0.2cm] && \hspace{-0.3cm}
=-\frac{~\as}{\rho}\int\limits_{0}^{\infty}{d\chi}\Big[\hat{A}(\cosh\chi)\notag
\sin\rho\chi\int\limits_{0}^{\infty}d{\rho'}\sin(\rho'\chi)\varphi_{0}(\rho',\chi')\Big]. \nonumber
\end{eqnarray}
Here the rapidity $\chi'$ parametrizes the relative 3-momentum, ${\bf\Delta}_{k',m'\lambda_{\mathcal{Q}}}$, and the total c.m. energy  $\sqrt{s}$,   Eq.~(\ref{s-uneq-Delta-k'}), as
\begin{eqnarray}
\label{chi-uneq} \boldsymbol{\Delta}_{k',m'\lambda_{\mathcal{Q}}}= m'\sinh\chi'{\bf
n}_{\Delta_{k',m'\lambda_{\mathcal{Q}}}},
|{\bf n}_{\Delta_{k',m'\lambda_{\mathcal{Q}}}}|=1,\\
\sqrt{s}=\frac{m'}{\mu}\Delta_{k',m'\lambda_{\mathcal{Q}}}^0,
\Delta_{k',m'\lambda_{\mathcal{Q}}}^0=m'\cosh\chi',\\
\Delta_{k',m'\lambda_{\mathcal{Q}}}^{02}-\boldsymbol{\Delta}_{k',m'\lambda_{\mathcal{Q}}}^2=m'^2,
\end{eqnarray}
$\rho=r/\lambda'$, and $\lambda'=1/m'$ is the Compton wavelength associated with the effective
relativistic particle of mass $m'=\sqrt{m_1m_2}$, the operator $\hat{A}$ is given by the expression
\begin{eqnarray}
\label{function-A-uneq-Delta-spin} \hat{A}\left(\cosh\chi'\right)=
\frac{1}{4}\left[a'\cosh^2\chi'+b'\right],
\end{eqnarray}
where
\begin{eqnarray}
\label{parameter-a-b-uneq-spin} \hspace{-0.2cm} a'=\left\{\begin{array}{ll} ~ g'^2 &
~ \text{\rm for}\,\hat{O}=\gamma_5\,\text{(\rm pseudoscalar)},\\[0.2cm]
~\frac{1}{2}\;g'^2 & ~ \text{\rm for}\,\hat{O}=\gamma_\mu\,\text{(\rm vector)},\\[0.2cm]
-\frac{1}{2}\;g'^2 & ~ \text{\rm for}\,\hat{O}=\gamma_5\gamma_\mu\,\text{(\rm pseudovector)};
\end{array}\right.\\[0.2cm]
\hspace{-0.6cm} b'=\left\{\begin{array}{ll} \displaystyle 1-g'^2 & \text{\rm for}\,\hat{O}=\gamma_5\,\text{(\rm
pseudoscalar)}
,\\[0.2cm]
\displaystyle\frac{3}{4}-\frac{1}{2}g'^2 & \text{\rm for}\,\hat{O}=\gamma_\mu\,\text{(\rm vector)}
,\\[0.2cm]
\displaystyle\frac{1}{4}+\frac{1}{2}g'^2 & \text{\rm for}\,\hat{O}=\gamma_5\gamma_\mu,\text{(\rm pseudovector)},
\end{array}\right.\notag
\end{eqnarray}
and the factor $g'$ is defined as
\begin{equation}
\label{factor-g} g'=\frac{m'}{2\mu}=\frac{m_1+m_2}{2\sqrt{m_1m_2}}\,.
\end{equation}
Obviously, the values of the parameters $a'$ and $b'$ in Eq.~(\ref{parameter-a-b-uneq-spin})
at $m_1=m_2=m$ coincide with the corresponding values  in Eq.~(\ref{parameter-a-b-eq-spin}).
It is worth   mentioning that  the rapidity $\chi'$, velocity $u$ and factor $g'$ are interrelated by the expression
\begin{equation}
\sinh\chi' = \frac{u}{g'\sqrt{1-u^2}}.
\end{equation}
The solution of Eq.~(\ref{eqvarphi-ell=0-int-uneq-rho-spin}) is sought in the form
\begin{eqnarray}
\label{form-solution-varphi-ell=0-int-uneq-rho-spin} \varphi_{0}(\rho,\chi')=\int\limits_{\alpha_{-}}^{\alpha_{+}}
d\zeta e^{i\rho\zeta}R_{0}(\zeta,\chi'),
\end{eqnarray}
where the integration is carried out in the complex plane along a contour between the end points $\alpha_{\pm}=-R\pm
i\varepsilon,R\rightarrow+\infty,\varepsilon\rightarrow+0$, the values $\pm\chi+2\pi ni\,~(n=0,~\pm1,~\ldots)$ are the
branch points of the function $R_{0}(\zeta,\chi)$, and the  integration contour  must not intersect the cuts which we take
from $-\infty+2\pi ni$ to $\pm\chi+2\pi ni$ (see Fig.~\ref{Fig-zeta-pl2}), that is,   just in the same manner  as before, c.f.~Refs.~\cite{Milton-Solovtsov_ModPL:01,SOP-Ch10,SOP-ChTMF-11,ChYAF2017}.

Substituting Eq.~(\ref{form-solution-varphi-ell=0-int-uneq-rho-spin}) into
Eq.~(\ref{eqvarphi-ell=0-int-uneq-rho-spin}), and  taking into account that
\[
\frac{1}{i\pi}\int\limits_{0}^{\infty}d\rho'\sin(\rho'\chi')e^{i\rho'\zeta}=
\frac{1}{i\pi}\frac{\chi'}{{\chi'}^2-{\zeta}^2}, ~~ \mbox{Im}\zeta>0,
\]
and performing   integration by parts, we arrive at the differential equation
\begin{eqnarray}
\label{eq-R-0-uneq-spin}
\frac{d}{d\zeta}\left[\left(\cosh\chi'-\cosh\zeta\right)R_0(\zeta,\chi')\right]-\\
-i\as\hat{A}(\cosh\zeta)R_0(\zeta,\chi')=0\notag
\end{eqnarray}
with the boundary condition
\begin{equation}
\label{condition-R-0-uneq-spin}
e^{i\rho\zeta}\left(\cosh\chi'-\cosh\zeta\right)R_0(\zeta,\chi')\Big|_{\zeta=\alpha_{-}}^{\zeta=\alpha_{+}}=0.
\end{equation}
Then the solution of Eq.~(\ref{eq-R-0-uneq-spin}) with such a boundary condition can be written as
\begin{eqnarray}
\label{solution-R-0-uneq-spin}
R_0(\zeta,\chi') =
 \frac{C_0(\chi')}{{(e^{\zeta}-e^{\chi'})^2}}\, {\displaystyle\exp\left[-\frac{i\as a'}{4}\sinh\zeta+(1-i\tilde{\rho}')\zeta+iB'\chi'\right]}\,
\left[\frac{e^{\zeta}-e^{-\chi'}}{e^{\zeta}-e^{\chi'}}\right]^{-1+iB'},
\end{eqnarray}%
where $C_{0}(\chi')$ is an arbitrary function of $\chi'$, the parameters $a',~b'$ are defined in
Eqs.~(\ref{parameter-a-b-uneq-spin}), and   $\tilde{\rho}'$ and $B'$ are
\begin{equation}
\label{B-ell=0-uneq-chi-spin} \tilde{\rho}'=\frac{\as a'\cosh\chi'}{4}\, , ~~~ B'=\frac{\as
(a'\cosh^2\chi'+b')}{4\sinh\chi'}\,.
\end{equation}

At $\chi'=i\kappa_n$, the parameter $B'$ is related to the quantization condition of the energy levels
\begin{equation*}
\label{condition-quantization-ell=0-uneq-kappa-spin} \frac{\as (a'\cos^2\kappa_n+b')}{4\sin\kappa_n}=n\,
\end{equation*}
with $~ n=1,~2,~\ldots$ and $\,0<\kappa_n<\pi/2\,$.

It should be emphasized that  for vanishing interaction, $\as\to0$, the asymptotics of the wave function
$\varphi_{0}(\rho,\chi')$ must reproduce the known free wave function,
\begin{equation}
\label{condition-varphi-ell=0-uneq-rho-spin} \lim_{\as\to 0}\varphi_{0}(\rho,\chi')\xrightarrow[\rho\to\infty]\,
\frac{\sin(\rho\chi')}{\sinh\chi'}.
\end{equation}

Substituting the solution~(\ref{solution-R-0-uneq-spin}) into Eq.~(\ref{form-solution-varphi-ell=0-int-uneq-rho-spin}) and
performing the $\zeta$-integration in the complex plane along the contour displayed in Fig.~\ref{Fig-zeta-pl2} with the end points $\alpha_{\pm}$ (c.f. Refs.~\cite{Milton-Solovtsov_ModPL:01,SOP-Ch10,SOP-ChTMF-11,ChYAF2017}), we obtain the resulting solution which does not contain the $i$-periodic constant:
\begin{eqnarray}&&
\label{solution-varphi-ell=0-uneq-rho-spin} \hspace{-0.3cm}\varphi_{0}(\rho,\chi')=2\,C_{0}(\chi')\,e^{iB'\chi'}
\sinh\left[\pi(\rho-\tilde{\rho}')\right]\\[0.2cm] &&
\times\int\limits_{-\infty}^{\infty}dx\frac{\displaystyle\exp\left[{\frac{i\as a'}{4}\sinh x+
\big(1+i(\rho-\tilde{\rho}')\big)\,x}\right]}{(e^{x}+e^{\chi'})^2}
\times\left[\frac{e^{x}+e^{-\chi'}}{e^{x}+e^{\chi'}}\right]^{-1+iB'},\notag
\end{eqnarray}
where  $C_{0}(\chi')$ is the normalization factor.
In addition, it is seen that  except for  the oscillating factor $\exp\left(i\as a'\sinh x/4\right)$  the wave function $\varphi_{0}(\rho,\chi')$
 coincides in form with the spinless case.  Moreover, for  $a'=0$ and $b'=2/g'$ Eq.~(\ref{solution-varphi-ell=0-uneq-rho-spin})  reproduces exactly the spinless wave function (see Refs.~\cite{SOP-Ch10,SOP-ChTMF-11}).
This circumstance  serves as a hint that we can approximate the wave function
~(\ref{solution-varphi-ell=0-uneq-rho-spin}) by setting  the oscillating factor $\exp\left(i\as a'\sinh x/4\right)$ equal to one; such an approximation  preserves all the symmetry characteristics of the solution
(\ref{solution-varphi-ell=0-uneq-rho-spin}) and  allows one to present the resulting
wave function (for the $s$-wave state) in an explicit analytical form. Indeed, by a proper change of
the integration variable $x$,
\begin{equation}
t=\frac{{\rm e}^{\chi'}}{ {\rm e}^{x'}+{\rm e}^{\chi'}  }, \qquad \chi' < \infty \label{change}
\end{equation}
  the
 integral in Eq.~(\ref{solution-varphi-ell=0-uneq-rho-spin}) leads to the well-known    representation
 for the hypergeometric function  $_2F_1  (a,b;c;z)$,
\begin{eqnarray}
_2F_1  (a,b;c;z)= \frac{\Gamma(c)}{\Gamma(b)\Gamma(c-b)} && \int\limits_0^1 dt t^{b-1} (1-t)^{c-b-1} (1-tz)^{-a}
\\[2mm] &&
Re\ c> Re\ b >0 ,\notag
\end{eqnarray}
exactly as in the case   of  spinors of equal masses (see Refs.~\cite{skachkov,ChYAF2019,CKS-2020}). The result is
\begin{gather}
\label{approximation-solution-varphi-ell=0-uneq-rho-spin}
\varphi_{0}(\rho,\chi')=2\pi C_{0}(\chi')e^{iB'\chi'-\chi'+i(\rho-\tilde{\rho}')\chi'}\times\\[0.2cm]
\times(\rho-\tilde{\rho}') _2F_1 \big(1-iB',1-i(\rho-\tilde{\rho}');2;1-e^{-2\chi'}\big),\notag
\end{gather}
where the parameters $\tilde{\rho}'$ and $B'$ are given in Eq.~(\ref{B-ell=0-uneq-chi-spin}) and the
normalization factor $2\pi C_{0}(\chi')$ is defined by the expression:
\begin{gather}
\label{normal-factor-varphi-ell=0-uneq-rho-spin} |2\pi C_{0}(\chi')|^2=e^{\pi B'}|\Gamma(1-iB')|^2,
\end{gather}
which   can be derived from the boundary condition~(\ref{condition-R-0-uneq-spin})
and the asymptotic behaviour~(\ref{condition-varphi-ell=0-uneq-rho-spin})
\begin{gather}
\label{approximation-varphi-ell=0-uneq-rho-spin}
\varphi_{0}(\rho,\chi')|_{\rho\gg 1}\approx\frac{2\pi C_{0}(\chi')e^{-\pi B'/2}}{\sinh\chi'|\Gamma(1-iB')|}
 \sin\big\{(\rho-\tilde{\rho}')\chi'+B'\ln\big[2(\rho-\tilde{\rho}')
\sinh\chi'\big]+\arg\Gamma(1-iB')\big\}.
\end{gather}

\subsection{Relativistic $\boldsymbol{S}$-factor}

Taking into account the relations~(\ref{BS-Psi-psi-RQP}) and
(\ref{determination-S-factor-eq}), the Coulomb  $S$-factor for a
system of two relativistic spinor particles of arbitrary masses is defined as
\begin{gather}
\label{determination-S-factor-uneq-spin}
\SuneqS (\chi')
=\lim_{\rho\to{i}}\left|e^{-\pi\tilde{\rho}'/2}\Gamma(1+i\tilde{\rho}')\;
\frac{\varphi_0(\rho,\chi')}{\rho}\right|^2 .
\end{gather}
The additional factor
$\exp({-\pi\tilde{\rho}'/2})\Gamma(1+i\tilde{\rho}')$  has been introduced to
ensure the  correct relativistic ($\chi'\gg 1$) behaviour of
$\SuneqS (\chi' \gg 1 )$ and, for $a'=0$ and $b'=2/g'$, to maintain  the spinless results. Recall that 
in the spinless case, the ultrarelativistic limit ($u\to 1$) of the S-factor, Eq.~(\ref{S-rel-uneq}),
is $S(u\to 1)=1$.
In the case of spinors, the hypergeometric function is strictly  defined for $u < 1$, c.f. Eq.~(\ref{change}), and consequently
for $u=1$ ($\chi' \to\infty$) the change of variables (\ref{change}) becomes ill defined. However,
the hypergeometric  function can be further defined at $u=1$ is such a way as to keep continuity with the previous results. This can be achieved by substituting in the hypergeometric function its argument $1-\exp(-2\chi')$ by unity and, at the same time, keeping the rest of the parameters as functions of $\chi'$. In this case, one can make use of the explicit expression of
$_2F_{1}(a,b;c,1)$
in terms of the Euler $\Gamma$-functions, $_2F_1(a,b;c,1)=\Gamma(c)\Gamma(c-a-b)/\Gamma(c-a)\Gamma(c-b);\ Re(c-a-b)>0$ and have the wave function (\ref{approximation-solution-varphi-ell=0-uneq-rho-spin}) defined in the whole interval of velocities, including $u=1$.

  Thus, we see that  the function
\begin{gather*}
\label{determination-physics-function-uneq-spin}
\psi_{0}(\rho,\chi')=e^{-\pi\tilde{\rho}'/2}\Gamma(1+i\tilde{\rho}')\varphi_{0}(\rho,\chi')
\end{gather*}
is the sought physical   spinor wave function for the Coulomb-like interaction~(\ref{Coulomb}).
Then, by making use
of  
relations~(\ref{approximation-solution-varphi-ell=0-uneq-rho-spin})--(\ref{determination-S-factor-uneq-spin}),
one obtains  the $S$-factor
for  two spinors  of arbitrary masses in the  form
\begin{equation}
\label{S-factor-uneq-spin}
\SuneqS (\chi')=\frac{\XuneqS(\chi')}{1-\exp\left[-\XuneqS(\chi')\right]}  
 e^{-\pi\tilde{\rho}'}\big|\Gamma(2+i\tilde{\rho}'){_2}F_1(1+iB',-i\tilde{\rho}';2;1-e^{-2\chi'})\big|^2
\end{equation}
with
\begin{equation}
\label{X-factor-chi-uneq-spin} \XuneqS (\chi')=2\pi B' =\frac{\pi\as(a'\cosh^2\chi'+b')}{2\sinh\chi'} \, ,
\end{equation}
where $\tilde{\rho}',~B'$ are given by Eq.~(\ref{B-ell=0-uneq-chi-spin}).
Quantities~(\ref{S-factor-uneq-spin}) and (\ref{X-factor-chi-uneq-spin})
can be expressed in terms of the velocity $u$, Eq.~(\ref{u-uneq}), and the relative velocity $u'_{\rm rel}$ of
an effective relativistic particle with mass $m'$:
\begin{equation}
\label{u'-rel-eff} u'_{\rm rel}=\frac{2u}{\sqrt{1-u^2}}.
\end{equation}
In particular, for Eq.~(\ref{X-factor-chi-uneq-spin}) one has
\begin{eqnarray}
\label{X-factor-u-uneq-spin} \XuneqS(u)=\frac{\pi\as \sqrt{1-u^2}}{2g'u}
\bigg[g'^2\left(a'+b'\right)
+\frac{a'u^2}{1-u^2}\bigg]=\frac{\pi\as}{g'u'_{\rm rel}}\left[g'^2\left(a'+b'\right)+\frac{a'}{4}u'^2_{\rm
rel}\right].
\end{eqnarray}

It is noteworthy that  the kinematic factor $g'$  defined by  
expression~(\ref{factor-g})  establishes the relationship between the
total c.m. energy $\sqrt{s}$ of the system of two particles
 of arbitrary masses $m_1$ and $m_2$ and the energy
$\Delta_{k',m'\lambda_{\mathcal{Q}}}^0$ of an effective particle with the
mass $m'$ and  3-momentum ${\bf\Delta}_{k',m'\lambda_{\mathcal{Q}}}$. Such an effective particle
 mimics the properties of the  two-body system under consideration with the same total c.m.
energy $\sqrt{s}$, see Eq.~(\ref{s-uneq-Delta-k'}).
Therefore, the kinematic factor $g'$ reflects the asymmetry between
these two systems. In addition, the factor~(\ref{factor-g}) reflects the
relation of the arithmetic mean of the masses of two
 particles to their geometric mean.   Consequently, $g' > 1$ for
$m_1\neq m_2$ and  $g' = 1$ for symmetric systems.

Equations  (\ref{S-factor-uneq-spin})-(\ref{X-factor-chi-uneq-spin}) define the most general expression for the relativistic $S$-factor of  systems of two spinor particles with arbitrary masses  obtained
within the quasi-potential approach~\cite{Kadyshevsky68,KadMS:72}.
Obviously, it includes all the  previously considered cases. So for $m_1\neq m_2$ and $a'=0$ and $b'=2/g'$    Eq.~(\ref{S-factor-uneq-spin}) reproduces exactly  the spinless $S$-factor~(\ref{S-rel-uneq}), while
for $m_1=m_2=m$ it coincides with the equal-masses $S$-factor~(\ref{S-factor-spin}).  The explicit analytical form of the $S$-factor, Eq.~(\ref{S-factor-uneq-spin}), allows for a detailed investigation of the spin-dependent Sommerfeld effect in the threshold-near region ($u\to 0$) where, as already mentioned,  one predicts an essential enhancement of the cross sections in processes with spinless particles. In this region the influence of spins on the scale of the effect
is of particular interest.   The relativistic limit is also of great interest  since, as  will be shown below, the $S$-factor turns out to be rather sensitive to the total spin  and   mass asymmetry of the system.

\begin{figure}[htbp]
 \includegraphics[width=0.6\textwidth]{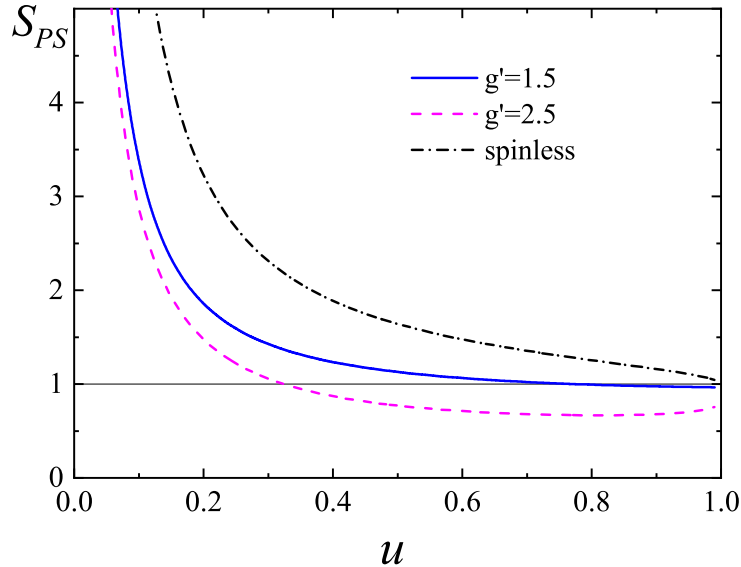}
\caption{(Color online) The behavior of the $S$-factor (\ref{S-factor-uneq-spin}) for the
pseudoscalar system, $a'=g'^2$ and $b'=1-g'^2$, as a function
of the velocity $u$, Eq.~(\ref{u-uneq}), at $\as=0.2$: solid
curve corresponds to $g'=1.5$, while the dashed one corresponds
to $g'=2.5$. The dot-dashed curve is for the spinless $S$-factor, Eq.~(\ref{S-rel-uneq}).}
\label{Fig-S_PS}
\end{figure}

\begin{figure}[htbp]
\includegraphics[width=0.6\textwidth]{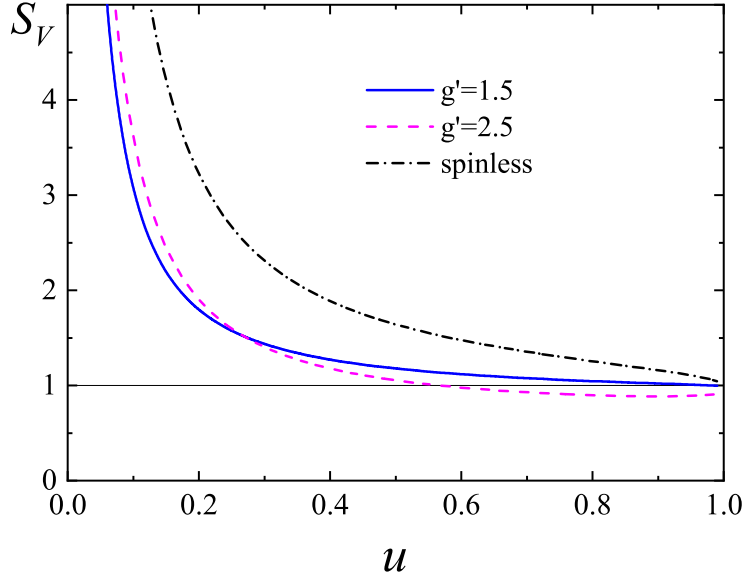}
\caption{(Color online)  The
same as in Fig.~\ref{Fig-S_PS} for the vector system, $a'=(1/2)g'^2$,
$b'=3/4-(1/2)g'^2$.} \label{Fig-S_V}
\end{figure}

\paragraph{Threshold-near region:}
Using the known properties of the hypergeometrical functions
\begin{equation*}
_2F_1(a,b,c,z/b)|_{b\to\infty}
\rightarrow \phantom{1}_1F_1(a,c,z)
\end{equation*}
 and
the known representation of the modulus of the $\Gamma$ functions,
\begin{eqnarray}&&
\left| \Gamma(x+iy)\right| = \left|\Gamma(x)\right| \prod_{n=1}^\infty\left(1+\frac{y^2}{(x+n)^2}\right)^{-1/2},
\qquad or\\ &&
\left| \Gamma(x+iy)\right|^2 = \left|\Gamma(x)\right|^2 \frac{\pi y}{\sinh(\pi y)}\prod_{n=1}^\infty
\left(\frac{1+ y^2/n^2}{1+y^2/(x+n)^2}\right),
\end{eqnarray}
one finds the $S$-factor at the threshold $u\to 0$ ($\chi'\rightarrow+0$) to be

\begin{eqnarray}
\label{lim-nr-S-factor-uneq-spin}
\SuneqS (\chi')\big|_{\chi'\rightarrow+0}  
\approx\frac{\pi\tilde{\alpha}_s(a'+b')/2\sinh\chi'}{1-\exp[-\pi\tilde{\alpha}_s(a'+b')/2\sinh\chi']}
\times\nonumber\\[2mm]
\times\frac{(\pi\tilde{\alpha}_sa'/4) \exp(-\pi\tilde{\alpha}_{s}a'/4)}{\sinh(\pi\tilde{\alpha}_{s}a'/4)}
\left(1+\frac{\tilde{\alpha}_s^2a'^2}{16}\right)
\left |{_1}F_1(-i\tilde{\alpha}_{s}a'/4;\,2;\,i\tilde{\alpha}_s(a'+b')/2)\right|^2.
\end{eqnarray}
\paragraph{Relativistic region:}
With the adopted definition of the wave function at $u=1$ ($\chi'\to \infty$),
 \begin{equation}\left .
 {_2}F_1(1+iB'(\chi'),-i\tilde{\rho}'(\chi');2;1-e^{-2\chi'})\right|_{\chi'\to \infty} \approx
  {_2}F_1(1+iB'(\chi'),-i\tilde{\rho}'(\chi');2;1)
 \end{equation}
  the relativistic limit, $\chi' \gg 1$, of the $S$-factor  is

\begin{equation}
\label{S-large}
\SuneqS (\chi')\big|_{\chi'\gg1}
\approx\frac{2\pi(B'-\tilde{\rho}')}{1-\exp[-2\pi(B'-\tilde{\rho}')]}\big|_{\chi'\gg1}
\approx1+\frac{\pi\as }{4}(a'+2b')e^{-\chi'}.
\end{equation}
From Eq.~(\ref{S-large}) and Eqs.~(\ref{parameter-a-b-uneq-spin})-(\ref{parameterbprime}) it follows that in the pseudoscalar case ($\hat{O}=\gamma_5$) for $g'>\sqrt{2}$ and in the vector one
($\hat{O}=\gamma_\mu$)  for $g'>\sqrt{3}$, the $S$-factor~(\ref{S-factor-uneq-spin}) in the relativistic limit tends to unity from
below, see also Figs.~\ref{Fig-S_PS} and \ref{Fig-S_V}.

\begin{figure}[htbp]
 \includegraphics[width=0.6\textwidth]{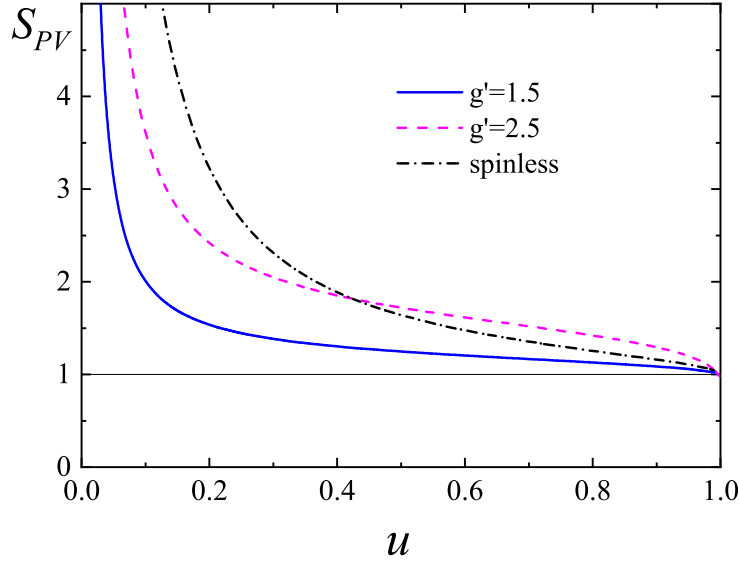}
 \caption{(Color online)  The same as in
Fig.~\ref{Fig-S_PS} for the pseudovector system,
$a'=-(1/2)g'^2$ and $b'=1/4+(1/2)g'^2$.} \label{Fig-S_PV}
\end{figure}

\section{Results}

We investigated numerically the $S$-factor for scalar, pseudo-scalar, vector and pseudovector two-particle
systems with arbitrary masses by the exact formula~(\ref{S-factor-uneq-spin}).
The results of   calculations of the $S$-factor as a function of the velocity $u$,   Eq.~(\ref{u-uneq}),
 are presented in Figs.~\ref{Fig-S_PS},~\ref{Fig-S_V} and~\ref{Fig-S_PV} for the pseudo-scalar, vector and pseudovector systems,
 respectively, for two values of the parameter $g'$. The solid curves corresponds to $g'=1.5$, while the dashed ones corresponds to
 $g'=2.5$. The dot-dashed curve is for the spinless $S$-factor, Eq.~(\ref{S-rel-uneq}). For definiteness, numerical calculations  have
 been performed for  $\tilde{\alpha}_s=0.2$. However, the general features of the $S$-factor do not depend on the concrete choice of
 $\tilde{\alpha}_s$.
From these figures one infers  that the  results  are rather sensitive to the value of $g'$  and, consequently,
to the parameters $a'$~and~$b'$.

\begin{figure}
 \includegraphics[width=0.6\textwidth]{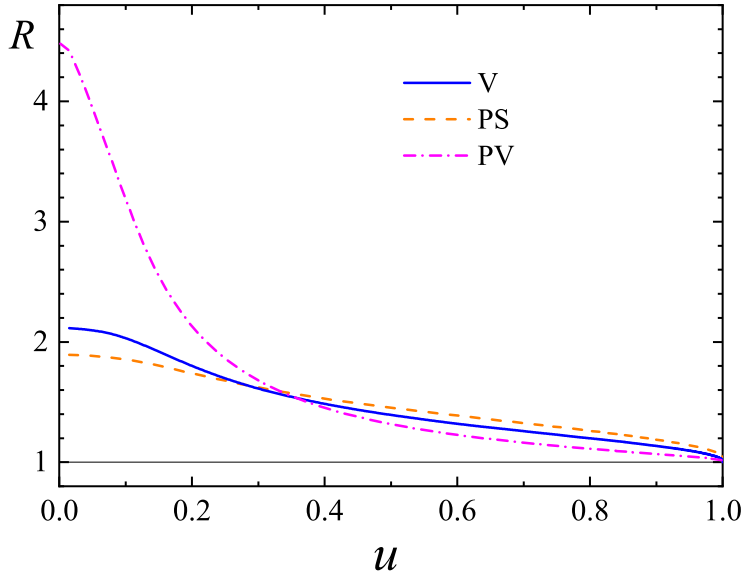}\vspace*{3mm}
 \caption{(Color online)  The ratio of
the $S$-factors of spinless systems to the spinor factors  as a function of the velocity $u$:
solid curve corresponds to  vector, dashed to pseudo-scalar and dot-dashed
curve to  pseudo-vector systems. The calculations have been performed for the same
values of the parameters as in Figs.~\ref{Fig-S_PS}-\ref{Fig-S_PV} for $g'=1.5$.}
 \label{ratio}
\end{figure}

As is seen from Figs.~\ref{Fig-S_PS}--\ref{Fig-S_PV} at low and moderate velocities,  the spinor $S$-factors are always
smaller than the $S$-factor for spinless particles. Note that, to our knowledge,
spin effects of spins in the relativistic $S$-factor have not yet been 
investigated. Therefore, a systematic   comparison of spinless and spinor factors can be considered as 
the very first results of the study of
spin effects in the relativistic $S$-factor. For a better illustration  of  spin effects,   we present in Fig.~\ref{ratio} the ratio of  spinless to spinor $S$-factors for  vector (solid), pseudo-vector (dash-dotted) and pseudo-scalar (dashed curve) systems.
It is seen that the effects of spins significantly reduce the Sommerfeld effect at the threshold in comparison with the spinless case,
while  at relativistic velocities they become insignificant and all, spinor and spinless factors, are practically the same and close
to unity.

Now  let us focus on  the regions of moderate and sufficiently large velocities, $u \gtrsim 0.5$.
Figures~\ref{Fig-S_PS} and \ref{Fig-S_V} demonstrate that, according to Eqs.~(\ref{S-large})
  and ~(\ref{parameter-a-b-uneq-spin})-(\ref{parameterbprime}), the $S$-factor of pseudo-scalar and vector systems   becomes less than unity and approaches its asymptotics,  $S\to 1$, from below. Such an unexpected
decrease of the $S$-factor, which reduces the cross sections, can have essential  impacts in treating the measured experimental data  at large $u$. Figure \ref{Fig-S_PV} shows that the $S$-factor for a pseudovector system is always larger than unity, despite the fact that  its
ultrarelativistic limit is also unity. The common limit for different systems can be qualitatively understood if one considers the case of (at least one) super-light particles, $m'\to 0$. Since,  as is well known, the $S$-factor characterizes the effects of the final (FSI)
or initial (ISI) state interaction, complementing the main process, see e.g. \cite{YoonWong:0005}, at ultrarelativistic velocities the
FSI/ISI effects vanish and the main reactions are determined solely by the corresponding Feynman diagrams without any additional
factors.

It should be stressed  that  since in our approach both  the argument ($r\equiv|\bf
r|$) of the Coulomb-like potential~(\ref{Coulomb}) and the relative
velocity $\vr$ are relativistic invariants~\cite{KadMS:72}, the
$S$-factor~(\ref{S-factor-uneq-spin}) is manifestly
relativistic invariant too. Therefore, due to the relation
$\vr=2u/(1+u^2)$, the velocity $u$, Eq.~(\ref{u-uneq}), and the
velocity $u'_{\rm rel}$ of the effective particle,
Eq.~(\ref{u'-rel-eff}), are also relativisic
invariants. Thus, instead of the previously used relative
velocity $\vr$, an appropriate parameter in the
$S$-factor~(\ref{S-factor-uneq-spin}) is the
velocity~(\ref{u'-rel-eff}) of an effective particle, which
has the same properties as the considered two-particle system.

\section{Conclusion}

In this paper, we have presented new calculations of the
relativistic $S$-factor, Eq.~(\ref{S-factor-uneq-spin}), within the covariant
 quasi-potential approach in the three-dimensional relativistic configuration
representation.    For the first time, the most general expression for the manifestly relativistic $S$-factor
of two-spinor systems with arbitrary masses has been derived explicitly. It is asserted  that the
relevant  velocity parameter of the approach is the one
defined by Eq.~(\ref{u'-rel-eff})  instead of the relative velocity  ${\vr}$  in so far commonly adopted in the literature. The spinless case can be obtained from the relativistic spinor $S$-factor by setting in Eqs.~(\ref{S-factor-uneq-spin})--(\ref{X-factor-chi-uneq-spin}) the spin
parameters $a'=0$ and $b'=4\sqrt{m_1 m_2}/(m_1+m_2)=2/g'$.
A comparison of the spinor factors with the spinless case persuades us that the spin effects play a significant role in the Sommerfeld effect making it, at small and moderate velocities, systematically smaller than the $S$-factor for  spinless systems. Thus, one can predict a decrease of the cross sections in a large kinematic  range of the relative velocity, the effect being amplified in the threshold-near region, see Fig.~\ref{ratio}.
 However, at sizeably large  velocities, $u\gtrsim 0.4$, the pseudo-vector   and the spinless $S$-factors become compatible to each other; moreover,   the pseudo-vector factor at large enough mass asymmetries ($g'\gtrsim 2.5$) becomes even larger than the spinless one,~c.f.   Fig.~\ref{Fig-S_PV}.  Another interesting observation is that at large velocities, $u\gtrsim 0.4-0.5$, the
 pseudo-scalar and vector factors for certain asymmetric systems
   cross the unity from above and approach the ultrarelativistic limit from below.
  Since in this region the cross sections become smaller in comparison with the main
  cross section determined by the corresponding Feynman diagrams, this circumstance can be, in some sense, referred to as the "anti"-Sommerfeld effect, contrarily to the "true" Sommerfeld effect discussed above~\footnote{Recall that  in the present paper we consider the attractive Coulumb-like potentials.}.
The "anti"-Sommerfeld effect  increases with   the mass asymmetry, $g'$, of the system and can cause
new impacts in the cross section of processes with highly asymmetric relativistic particles.

An interrelation  of the new $S$-factor~(\ref{S-factor-uneq-spin})
with the  previously  considered  in the literature (symmetric spinors and
spinless two-body systems of arbitrary masses) has been settled. Our analysis of the obtained results at the kinematic limits of the relative velocity $u$, i.e. in the deep nonrelativistic and ultrarelativistic regions,
shows that the approach  reproduces  the well-known nonrelativistic behaviour
of the spinless  factor for the symmetric systems ($m_1=m_2=m$)   as well as
provides  the expected  correct ultrarelativistic limits for two-spinor systems.

  \section{Summary}
In summary, we present the most general expression for the relativistic Sommerfeld--Gamov--Sakharov $S$-factors   for spinless, pseudo-scalar, vector and pseudo-vector relativistic particles with arbitrary masses derived
within the covariant, relativistic  quasi-potential approach in the
3-dimensional configuration representation in Lobachevsky space.
 The new $S$-factor is thoroughly investigated in the whole kinematic  region of the relative velocity $u$, including detailed studies in the deep threshold region and   ultrarelativistic limit. The effects of spins and mass-asymmetry are discussed.

\section*{Acknowledgments}

The authors gratefully acknowledge helpful
discussions with S.~M.~Dorkin, V.~I.~Lashkevich and E.~A.~Tolkachev and their continuous interest in the research topic.
This work was supported in part  by the International Program of
Cooperation between the Republic of Belarus and JINR.

\end{document}